\def\rfr#1{eq. (\ref{#1})}
\def\rfrs#1#2{eq. (\ref{#1})-eq. (\ref{#2})}

\def\derp#1#2{\rp{\partial{#1}}{\partial{#2}}}

\def\eqi{\begin{equation}}
\def\eqf{\end{equation}}
\def\eqia{\begin{eqnarray}}
\def\eqfa{\end{eqnarray}}
\def\Om{\mathit{\Omega}}
\def\rp#1#2{{#1\over#2}}
\def\lb#1{\label{#1}}

\def\ton#1{\left(#1\right)}
\def\qua#1{\left[#1\right]}

\documentclass[article]{elsart}
\usepackage{natbib}
\usepackage{lineno,hyperref}
\modulolinenumbers[5]
\usepackage{url}
\usepackage{float}
\usepackage{amsmath,textgreek,w-greek}
\usepackage{amscd,lineno}
\usepackage{amssymb}
\usepackage{graphicx,epsfig}
\RequirePackage{color}
\begin{document}
\begin{frontmatter}

\title{\textcolor{black}{The impact of the orbital decay of the LAGEOS satellites on the frame-dragging tests}}

\author{Lorenzo Iorio\thanksref{footnote2}}
\address{Ministero dell'Istruzione, dell'Universit\`{a} e della Ricerca (M.I.U.R.)-Istruzione\\ Fellow of the Royal Astronomical Society (F.R.A.S.)}
\thanks[footnote2]{Address for correspondence:  Viale Unit\`{a} di Italia 68, 70125, Bari (BA), Italy}
\ead{lorenzo.iorio@libero.it}
\ead[url]{http://digilander.libero.it/lorri/homepage$\_$of$\_$lorenzo$\_$iorio.htm}

\begin{abstract}
The laser-tracked geodetic satellites LAGEOS, LAGEOS II and LARES are currently employed, among other things, to measure the general relativistic Lense-Thirring effect in the gravitomagnetic field of the spinning Earth with the hope of providing a more accurate test of such a prediction of the Einstein's theory of gravitation than the existing ones. The secular decay $\dot a$ of the semimajor axes $a$ of such spacecrafts, recently measured in an independent way to a $\sigma_{\dot a}\approx 0.1-0.01$ m yr$^{-1}$ accuracy level, may indirectly impact the proposed relativistic experiment through its connection with the classical orbital precessions induced by the Earth's oblateness $J_2$. \textcolor{black}{Indeed,} the systematic bias due to the current measurement errors $\sigma_{\dot a}$ is of the same order of magnitude of, or even larger than,  the expected relativistic signal itself; moreover, it grows linearly with the time span $T$ of the analysis. \textcolor{black}{Therefore, the parameter-fitting algorithms must be properly updated in order to suitably cope with such a new source of systematic uncertainty. Otherwise,} an improvement of one-two orders of magnitude in measuring the orbital decay of the satellites of the LAGEOS family would be required to reduce this source of systematic uncertainty to a percent fraction of the Lense-Thirring signature.
\end{abstract}


%


\end{frontmatter}

\parindent=0.5 cm

\section{Introduction}
The so-called general relativistic Lense-Thirring effect \citep{LT18} consists of small secular orbital precessions affecting the motion of a test particle in geodesic motion about a central rotating body. They are driven by the gravitomagnetic\footnote{Such a denomination comes from the purely formal resemblance of the linearized Einstein field equations in the weak-field and slow-motion approximation with the Maxwell equations of electromagnetism.} component of the gravitational field of the spining mass due its proper angular momentum. As far as the Earth is concerned, the Satellite Laser Ranging (SLR) technique has been used so far in conjunction with the geodetic satellites LAGEOS and LAGEOS II to attempt to measure such a manifestation of frame-dragging\footnote{The GP-B mission \citep{2011PhRvL.106v1101E} measured another gravitomagnetic effect in the field of the Earth.} \textcolor{black}{\citep{2004Natur.431..958C, 2010ASSL..367..371C}}. A third member of this class of passive, spherical laser targets, named LARES, was launched  a few years ago \citep{2013AcAau..91..313P}, and it will be used to try to refine the existing tests in view of its improved manufacturing and design which will reduce the direct impact of the non-gravitational perturbations; according to its proponents, it should be possible to reach a $\approx 1\%$ total accuracy \citep{2013NuPhS.243..180C}.
For recent overviews of the performed and ongoing attempts to detect the Lense-Thirring effect with artificial Earth's satellites and related discussions on the realistic accuracy level reached, see, e.g., \citet{2011Ap&SS.331..351I}, \textcolor{black}{\citet{2012NewA...17..341C}}, \citet{2013OPhy...11..531R, 2013AcAau..91..141I, 2013NuPhS.243..180C, 2015AcAau.113..164R} and references therein.

In this paper, we want to address a further source of potential systematic uncertainty in the total error budget of the ongoing relativity experiment involving the three satellites of the LAGEOS family. Basically, the Newtonian  precession of the orbital plane induced by the asphericity of the Earth \citep{2008orbi.book.....X} is nominally several orders of magnitude larger than the general relativistic gravitomagnetic one. Such a classical feature, which acts as a major systematic error, depends, among other factors, on some of the satellite's orbital parameters; thus, long-term variations in them indirectly affect also the orbital precession itself inducing a further bias which has to be properly assessed and subtracted from the signal from which the Lense-Thirring effect should be extracted.
In particular, it turns out that the size of the orbits of the satellites considered is steadily diminishing over the years \citep{Sosproc, SosBOOK}. Although the gravitomagnetic field does not directly affect such an orbital feature, the resulting additional shift in the orbital plane may alias the corresponding Lense-Thirring precession by \textcolor{black}{potentially} corrupting the overall accuracy level in this frame-dragging test \textcolor{black}{in absence of appropriate countermeasures in the data reduction procedure}.

The relevant orbital parameters of LAGEOS, LAGEOS II, LARES are summarized in Table \ref{tavola1}\textcolor{black}{, in which the following values of the key fundamental constants and geophysical parameters of the Earth were adopted: $G=66,451.1$ kg$^{-1}$ m$^3$ yr$^{-2}$, $c = 9.46073\times 10^{15}$ m yr$^{-1}$, $GM = 3.96959\times 10^{29}$ m$^3$ yr$^{-2}$, $S = 1.84928\times 10^{41}$ kg m$^2$ yr$^{-2}$, $J_2 = 0.00108262$, $R$ = 6,378,136.6 m.
}
\begin{table}[H]
\centering
\caption{Relevant orbital parameters of the satellites of the LAGEOS family: $a$ is the semimajor axis, $e$ is the eccentricity, $I$ is the inclination to the Earth's equator, $n$ is the Keplerian mean motion.}
\label{tavola1}
\begin{tabular}{llll}
\noalign{\smallskip}
\hline
 & LAGEOS & LAGEOS II & LARES \\
\hline
$a$ (km) & $12,274$ & $12,158$ & $7,820$\\
$e$  & $0.0039$ & $0.0137$ & $0.0007$\\
$I$ (deg) & $109.90$ & $52.67$ & $69.50$\\
$n$ (yr$^{-1}$) & $1.465\times 10^4$ & $1.486\times 10^4$ & $2.881\times 10^4$\\
\hline
\end{tabular}
\end{table}
\section{Evaluating the bias due to the satellite's orbital decay}
\subsection{The classical and relativistic orbital precessions}
In the following, a coordinate system whose $z$ axis is aligned with the Earth's spin axis will be assumed.

For a given value of its inclination $I$ to the Earth's equator, the location of the satellite's orbital plane in the inertial space is determined by the longitude of the ascending node $\Om$: it is in angle in the equatorial plane reckoned from a reference $x$ direction to the line of the nodes which, in turn, is determined by the intersection of the orbital plane with the equatorial plane itself. If the primary was pointlike or perfectly spherical, the orbital plane would remain fixed in space, so that $\Om$ would keep a constant value. Actually, every astronomical body such as our planet does rotate, so that its shape is distorted by the centrifugal effects which make its gravitational field to depart from spherical symmetry. To the Newtonian level, the gravitational potential is usually expanded in multipolar coefficients; among them, the even zonal ones $J_{\ell},~\ell=2,4,6,\ldots$ are of particular importance since they induce orbital perturbations which do not vanish when averaged out over one orbital revolution.
In particular, the satellite's node rate induced by the first even zonal $J_2$ of the central body is \citep{2008orbi.book.....X}
\eqi
\dot\Om_{J_2} = -\rp{3}{2}n\ton{\rp{R}{a}}^2\rp{\cos I J_2}{\ton{1-e^2}^2},\lb{dOdt}
\eqf
where the Keplerian mean motion is
\eqi
n = \sqrt{\rp{GM}{a^3}}\lb{kepmean}.
\eqf
In \rfrs{dOdt}{kepmean}, $G$ is the Newtonian constant of gravitation, and $M,R,J_2$ are the mass, the equatorial radius and the dimensionless first even zonal harmonic of the primary, respectively; $a$ and $e$ are the satellite's semimajor axis and eccentricity determining its orbital size and shape, respectively.

The Lense-Thirring node rate is \citep{LT18}
\eqi
\dot\Om_{\rm LT} = \rp{2GS}{c^2 a^3\ton{1-e^2}^{3/2}},\lb{dLTdt}
\eqf
where $c$ is the speed of light in vacuum and $S$ is the proper angular momentum of the primary.

For the LAGEOS-type satellites, the Newtonian disturbing precessions due to $J_2$ and the other even zonals of higher degree are nominally several orders of magnitude larger than the relativistic ones; thus, although the geopotential is modelled in the softwares routinely used to analyze satellites' data, their mismodelled components due to the current uncertainties $\delta J_{\ell},~\ell=2,4,6,\ldots$ are still too large with respect to the Lense-Thirring precession of \rfr{dLTdt}. As a consequence, several strategies have been devised so far to circumvent such a potentially fatal drawback; for details, see, e.g., \citet{2011Ap&SS.331..351I, 2013OPhy...11..531R, 2013NuPhS.243..180C} and references therein.

As we will show in the next Section, the present-day level of uncertainty in our knowledge of the even zonals is not the only source of systematic bias connected with the classical node precession of \rfr{dOdt}.
\subsection{When the semimajor axis does secularly vary}
By allowing for a secular variation of the semimajor axis $a$
\eqi a(t) = a_0 +\dot a t,\lb{aditi} \eqf where $a_0$ is its value at some chosen reference epoch, the integrated shift of the node over a time span $T$ much longer\footnote{The precession of \rfr{dOdt} is averaged over a satellite's full orbital revolution.} than the orbital period is, from \rfr{dOdt},
\eqi\Delta\Om_{J_2}=\int_0^T \dot\Om_{J_2}(t)dt = \rp{3n_0 R^2\cos I J_2}{5a_0 \ton{1-e^2}^2\dot a}\qua{-1 + \ton{1 + \lambda}^{-5/2}}.\lb{shiftO}\eqf
In \rfr{shiftO}, we defined the dimensionless quantity
\eqi \lambda\doteq \rp{\dot a T}{a_0};\eqf moreover,
\eqi n_0 \doteq \sqrt{\rp{GM}{a_0^3}}.\eqf In the case of the LAGEOS-type satellites, it is
\eqi\left|\lambda\right|\ll 1\eqf over time intervals some years long since the measured semimajor axis' rates $\dot a$ are less than a meter per year \citep{Sosproc, SosBOOK}, while the nominal values of their semimajor axes are as large as thousands of kilometers, as shown by Table \ref{tavola1}.
Thus, in view of the fact that
\eqi
-1+\ton{1+\lambda}^{-5/2}\approx -\rp{5}{2}\lambda +\rp{35}{8}\lambda^2 + \mathcal{O}\ton{\lambda^3},
\eqf
\rfr{shiftO} can be approximately written as
\eqi
\Delta\Om_{J_2}\approx -\rp{3}{2}n_0\ton{\rp{R}{a_0}}^2\rp{\cos I J_2 }{\ton{1-e^2}^2}T + \rp{21 n_0 R^2\cos I J_2 \dot a}{8a_0^3\ton{1-e^2}^2}T^2 +\mathcal{O}\ton{\dot a^2 T^3}.\lb{eccolo}
\eqf

As far as the Lense-Thirring effect is concerned, by inserting \rfr{aditi} into \rfr{dLTdt}, the gravitomagnetic node shift over the time interval $T$ turns out to be
\eqi
\Delta\Om_{\rm LT} = \int_0^T\dot\Om_{\rm LT}(t) dt = \rp{2GS \ton{1 + \rp{\lambda}{2}}}{c^2 a_0^3\ton{1-e^2}^{3/2}\ton{1+\lambda}^2}T.\lb{esattaLT}
\eqf
Actually, \rfr{esattaLT} can be approximated as
\eqi
\Delta\Om_{\rm LT} \approx \rp{2GS}{c^2 a_0^3\ton{1-e^2}^{3/2}}T+\mathcal{O}\ton{\dot a T^2}\lb{shiftLT}
\eqf
since
\eqi
\rp{1+\rp{\lambda}{2}}{\ton{1+\lambda}^2}\approx 1 -\rp{3}{2}\lambda + \mathcal{O}\ton{\lambda^2}.
\eqf

In order to correctly evaluate the impact of our unavoidably imperfect knowledge of the secular rate of the satellite's semimajor axis on the Lense-Thirring shift of \rfr{shiftLT}, let us calculate
\eqi\delta\ton{\Delta\Om_{J_2}} \leq \left|\derp{\Delta\Om_{J_2}}{\dot a}\right|\delta\dot a,\lb{derivo}\eqf where $\delta\dot a$ may represent either the uncertainty in adequately modeling the rate of change of $a$ due to some proposed physical phenomenon or, if available, the experimental error in phenomenologically determining it from actual data reductions; in the latter case, for the sake of clarity, we will use the symbol $\sigma_{\dot a}$.

From \rfr{eccolo}, it turns out that, within the approximations used, the uncertainty in the $J_2-$induced node shift is quadratic in the time interval, amounting to
\eqi
\delta\ton{\Delta\Om_{J_2}} \lesssim \rp{21 n_0 R^2\left|\cos I\right| J_2 }{8a_0^3\ton{1-e^2}^2}T^2\delta\dot a.\lb{kazzo}
\eqf
Thus, from \rfr{kazzo} and \rfr{shiftLT} it is possible to calculate the following dimensionless parameter
\eqi\xi_{\Om}\doteq\rp{\delta\ton{\Delta\Om_{J_2}}}{\Delta\Om_{\rm LT}} \lesssim \rp{21 c^2 n_0 R^2\left|\cos I\right| J_2 }{16GS\ton{1-e^2}^{1/2}}T\delta\dot a;\lb{Xis}\eqf
it should be noticed that our measure of the impact of the semimajor axis secular rate on the Lense-Thirring test grows linearly with time.

An important feature of the following analysis consists of the fact that we will rely upon the phenomenologically determined values of the secular decays of the semimajor axes of the satellites of the LAGEOS family from real data reductions of long observational records \citep{Sosproc, SosBOOK} which, in the case of LAGEOS and LAGEOS II, amount to decades. In other words, we will not attempt to invoke some known or exotic physical phenomenon able to produce such an observed orbital feature in order to model it along with the associated mismodeling. \textcolor{black}{Several effects, mainly rooted in standard non-gravitational physics, have been proposed and discussed so far; see, e.g., \citet{1982CeMec..26..361R,1987JGR....92.1287R, 1988JGR....9313805R, 1991JGR....96..729S, 2015CQGra..32o5012L} and references therein.} In this respect, the uncertainties $\delta\dot a$ will be the errors $\sigma_{\dot a}$ released by independent analysts, neither involved with the LARES team nor with ourselves, who processed the SLR observations for scopes unconnected with testing frame-dragging \citep{Sosproc, SosBOOK}. Table \ref{tavola2} summarizes our findings for LAGEOS, LAGEOS II, LARES.
\begin{table*}[h]
\centering
\caption{The figures quoted for the decay $\dot a\pm \sigma_{\dot a}$ of the semimajor axis were phenomenologically determined in actual analyses of SLR observations \citep{Sosproc, SosBOOK}: as such, they do not come from any a priori physical modeling. The values for $\xi_{\Om}$ were obtained from \rfr{Xis} with the satellites' parameters of Table \ref{tavola1} for $T=17$ yr; thus, the value for LARES has to be considered as merely indicative since the quoted $\sigma_{\dot a}$ was determined by analyzing just a yearly data record. }
\label{tavola2}
\begin{tabular}{llll}
\noalign{\smallskip}
\hline
 & LAGEOS & LAGEOS II & LARES \\
\hline
$\dot a$ (m yr$^{-1}$) & $-0.203$ & $-0.239$ & $-0.775$ \\
$\sigma_{\dot a}$ (m yr$^{-1}$) & $0.035$ & $0.037$ & $0.14$ \\
$\xi_{\Om}$ ($\%$) & $125$ & $238$ & $\textcolor{black}{1011}$ \\
\hline
\end{tabular}
\end{table*}
It can be noticed that the systematic bias on the uncombined nodes of each satellite is at the $\approx 100\%$ level of the relativistic effect over the  temporal interval adopted.

As far as LARES is concerned, it must be stressed that the larger uncertainty $\sigma_{\dot a}$ with respect to LAGEOS and LAGEOS II might be due, at least to some extent, to the comparatively shorter data record analyzed  so far. Anyway, even by arguing that it could be finally brought down to the accuracy level of its  long-lived cousins over multidecadal time spans, its bias on the predicted gravitomagnetic \textcolor{black}{node shift will} remain significantly higher than a few percent. \textcolor{black}{Moreover, great care is needed in correctly evaluating possible future improvements in the determination of $\dot a$ to avoid unphysical conclusions. Indeed, the released errors $\sigma_{\dot a}$ do not necessarily reflect realistic accuracy in our knowledge of the orbital rates of decay, being, instead, just a sort of measure of the internal precision of the data reduction procedure implemented. Otherwise, one would be drawn to paradoxical conclusions such as that the satellite's orbit radius could be finally known with a better accuracy than, say, the Earth's radius itself, which is currently known at $0.1$ m level \citep{2010ITN....36....1P}. Indeed $a$ can  be naively thought as the length of a vector  pointing from the Earth's center of mass to the spacecraft which, in turn, is (approximatively) the sum of a vector from the Earth's center of mass to a SLR station and of a vector from the latter to the spacecraft: it would be absurd to assume that, over any time interval, the length of the sum  of such vectors could be known more accurately than any one of the vectors themselves to be summed.} \textcolor{black}{Finally, it is worthwhile noticing that, after more than 30 years, it does not seem that a breakthrough in the accuracy of the determination of the orbital decay rate of LAGEOS has occurred so far. Indeed, if, on the one hand, \citet{1982CeMec..26..361R} quotes a value accurate to $\sigma_{\dot a}=0.1$ mm d$^{-1} = 0.036$ m yr$^{-1}$, on the other hand,  \citet{Sosproc, SosBOOK} still report $\sigma_{\dot a}=0.035$ m yr$^{-1}$  after 30 years or so (see Table \ref{tavola2}). }

\begin{figure}[h]
\centering
\centerline{
\vbox{
\begin{tabular}{cc}
\epsfysize= 10.0 cm\epsfbox{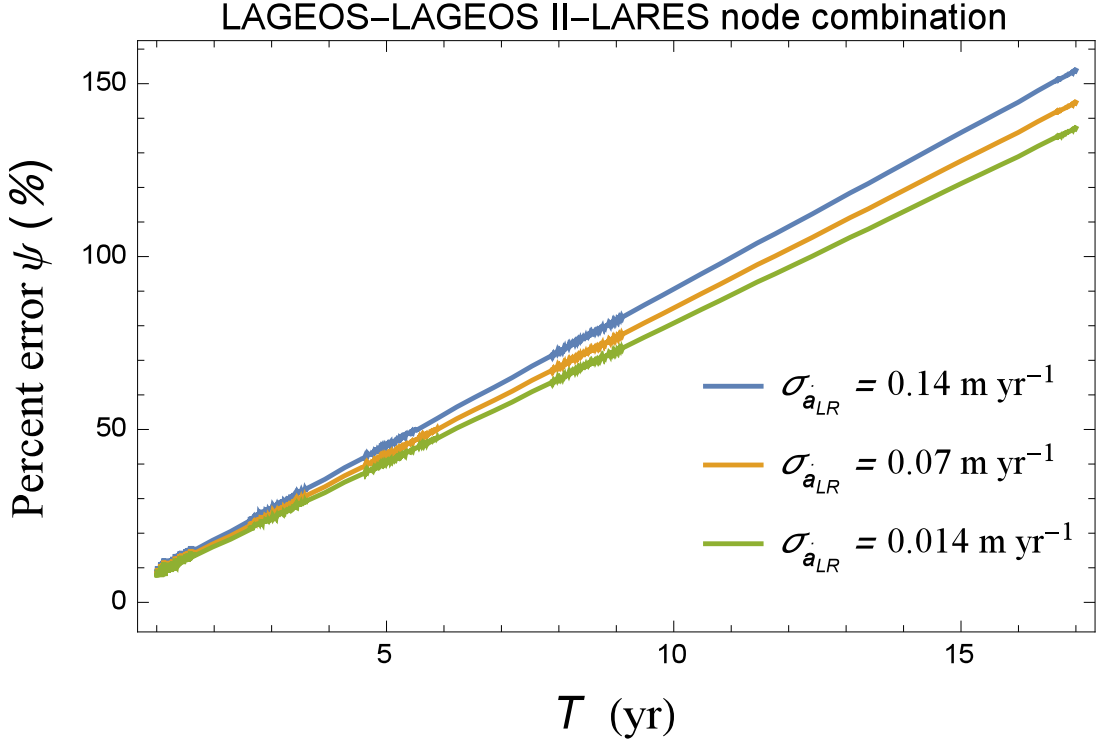} \\
\end{tabular}
}
}
\caption{Plot of \rfr{psi} as a function of the time span $T$ for different values of $\sigma_{\dot a_{\rm LR}}$. The errors $\sigma_{\dot a_{\rm L}},~\sigma_{\dot a_{\rm LII}}$ for LAGEOS and LAGEOS II were kept fixed to the values quoted in Table \ref{tavola2}. In calculating \rfr{psi}, the exact expressions for $\delta\ton{\Delta\Om_{J_2}}$ of \rfr{derivo} with \textcolor{black}{\rfr{kazzo}} and $\Delta\Om_{\rm LT}$ of \rfr{esattaLT} were assumed, i.e. no a priori simplifying assumptions on $\lambda$ were considered. }\label{figura}
\end{figure}

However, the actual attempts to measure the Lense-Thirring effect will be implemented by linearly combining the nodes of LAGEOS, LAGEOS II, and LARES by means of suitable coefficients purposely designed to remove, in principle, the mismodeling in the first two even zonals of the geopotential. More precisely, the following time series \textcolor{black}{will be analyzed}
\eqi
f^{\rm 3L} = \Delta\Om^{\rm LAGEOS} +c_1\Delta\Om^{\rm LAGEOS~II}+c_2\Delta\Om^{\rm LARES},\lb{combo}
\eqf
with
\begin{align}
c_1 & = 0.3586, \\ \nonumber \\
c_2 & = 0.0751.
\end{align}
Thus, we must apply our strategy to \rfr{combo} by assessing the impact of the observed decays of the LAGEOSs' semimajor axes on the combination of the Lense-Thirring node shifts, i.e., we have to study
\eqi \psi\doteq\rp{\delta\ton{\Delta f_{J_2}^{\rm 3L}}}{\Delta f^{\rm 3L}_{\rm LT}}.\lb{psi}\eqf
In Figure \ref{figura}, we plot \rfr{psi} as a function of the time span for different \textcolor{black}{constant} values of $\sigma_{\dot a_{\rm LR}}$ in order to consider possible amelioration of the accuracy of the determination of the LARES semimajor axis decay as more data will be collected and processed. \textcolor{black}{It turns out that, even by allowing for a linearly improving accuracy in the orbital decay of LARES over the years in such a way that it gets 10 times better after 10 years, no noticeable modifications in the pattern depicted occurs.}
It turns out that, even on relatively short temporal intervals and by assuming a potential improvement of one order of magnitude in measuring the orbital decay of LARES, the bias due to $\dot a$ is quite significant.

\textcolor{black}{Finally, one may wonder if the second even zonal harmonic $J_4$ of the geopotential may be of some concern, having some non-negligible impact on the Lense-Thirring test. Actually, by repeating the same reasoning as for $J_2$, it turns out that the resulting systematic bias $\delta\ton{\Delta f^{\rm 3L}_{J_4}}/\Delta f^{\rm 3L}_{\rm LT}$ is below the percent level over decadal time spans.}
%
\section{Summary and conclusions}
Although the satellite's semimajor axis $a$ is not directly affected by the gravitomagnetic field of a rotating primary-which, instead, chenges the orbital node $\Om$-its secular variation $\dot a$ has an indirect impact on the latter one through the huge classical precession $\dot \Om_{J_2}$ caused by the primary's oblateness $J_2$. As such, it must be included in the sources of systematic biases of the overall error budget of the ongoing space-based experiment aimed to measure the Lense-Thirring effect in the Earth's gravitational field by means of a linear combination of the nodes of LAGEOS, LAGEOS II, LARES.

Extended analyses of SLR observations have recently proven that, at present, the semimajor axes of such spacecrafts secularly decay at rates of the order of about $0.2-0.7$ m yr$^{-1}$, determined to a $\approx 0.01-0.1$ m yr$^{-1}$ accuracy level. As a consequence, the resulting residual node precessions due to $J_2$ induce a systematic uncertainty in the general relativistic signatures which grows linearly in time, reaching a size comparable to, or even larger than, the Lense-Thirring effect itself.

It must be stressed that our analysis is not based on the modeling of some possible known or exotic physical mechanism causing the aforementioned orbital decay. Instead, it relies upon phenomenological determinations of $\dot a$ from actual data reductions performed independently by other teams of SLR analysts, not involved in the relativity experiment. In other words, we have used only measured quantities, not calculated ones on the basis of some model.

Furthermore, it would be incorrect to claim that, since the geopotential is actually modeled in the data reduction softwares commonly used, one should, instead, consider only the mixed effect of the errors in either $\dot a$ and $J_2$. Indeed, as elementary error theory shows, when some quantity $g\left(q_i\right)$ depends on several parameters $q_i,~i=1,2,\ldots N$ affected by experimental/observational errors $\sigma_{q_i},~i=1,2,\ldots N$, the upper limit of the uncertainty in it is calculated by taking the sum of the absolute values of the derivatives with respect to each of the parameters scaled by the errors in them, i.e. $\delta g\leq \sum_{i=1}^N \left|\partial_{q_i} g\right|\sigma_{q_i}$. Additional mixed terms proportional to $\sigma_{q_i} \sigma_{q_j}$ can occur when correlations among the various parameters are present; it is not the case here since  both $\dot a$ and $J_2$ are determined in a completely independent way from different data sets of different satellites.

It is also useful to remind that one should not confuse the statistical error arising from data analysis with the assessment  of the overall systematic uncertainty, which is based on independent sensitivity analyses recurring also to several sources of information. As such, it would be incorrect to argue that a proposed source of systematic bias cannot exist just because the empirical time series apparently does not show it. Suffice it to recall the case of the earlier tests involving the nodes of LAGEOS and LAGEOS II and the perigee of LAGEOS II, when it was later shown \citep{Riesetal03, 2003CeMDA..86..277I} that the overall systematic uncertainty was likely as large as $50-100\%$, if not larger, of the expected Lense-Thirring signal despite the fact that such huge biases were absent from the time series of the combined residuals of the orbital elements usually showed, which was characterized by a relatively small statistical error (see, e.g., Section 3.1 of \citet{2013OPhy...11..531R} and references therein).

In conclusion, we have identified a further, non-negligible component of the total error budget in the LAGEOS-LARES relativity experiment whose magnitude is, at present, comparable with the sought Lense-Thirring signal itself. \textcolor{black}{Thus, the statistical data analysis should take into account of it, and the data reduction algorithms must be properly updated. Otherwise,} an improvement of one-two orders of magnitude in the measurement of the orbital secular decay of the satellites involved would be needed to reduce it to a percent level.
\textcolor{black}{\section*{Acknowledgements}
I gratefully thank four anonymous reviewers and the Associate Editor whose patience, constructive suggestions and criticisms remarkably contributed to improve the manuscript.}

\bibliography{semimabib}{}

\end{document}